# Blind source separation using Fast-ICA with a novel nonlinear function

Pengfei Xu[a,*], Yinjie Jia[a,b] and Zhijian Wang[a]

[a]College of Computer and Information, Hohai University, Nanjing, Jiangsu 210098, China
[b]Faculty of Electronic Information Engineering, Huaiyin Institute of Technology, Huaian, Jiangsu 223003, China

**Abstract:** Blind source separation(BSS) is a hotspot in signal processing, and independent component analysis (ICA) is a very effective tool for solving the BSS problem. In order to improve the performance of the separation, a new nonlinear function 'sin' was introduced. It can replace the commonly used classical functions ('tanh', 'gauss' and 'pow3') and does not need to select different nonlinear functions according to the Gauss property of signals. The two Matlab simulation results show that the improved Fast-ICA algorithm with the proposed nonlinearity can not only improve the separation accuracy but also speed up the convergence of blind source separation.

## 1  Introduction

The seminal work on blind source separation is by Jutten and Herault [1] in 1985, the problem is to extract the underlying source signals from a set of linear mixtures, where the mixing matrix is unknown. That means BSS seeks to recover or separate original source signals from their mixtures without any prior information on the sources or the parameters of the mixtures. Independent Component Analysis is the commonest and basic method to solve BSS problem, and it has been widely used in speech and image signal processing. Jutten, Herault, Comon and Sorouchyari published three classical papers on blind source separation(an adaptive algorithm based on neuromimetic architecture, problems statement and stability analysis) in 1991 and proposed the concept of ICA, which made the research of BSS developed greatly. Blind source separation of instantaneous mixed signals is systematically analyzed, which provides a more accurate explanation for independent component analysis[2]. Hyvarinen proposed a fixed-point training algorithm based on the non-Gaussian nature of the source signal. He first proposed Fast-ICA algorithm based on kurtosis [3] in 1997, and then proposed Fast-ICA algorithm based on negative entropy[4] in 1999. In recent years, some progress has been made in the research on the selection of non-linear functions in Fast-ICA algorithm[5-8].

In this paper, we propose one new non-linear function to replace the commonly used nonlinearities in the Fast-ICA to improve the convergence property. The rest of this paper is organized as follows. The improved Fast-ICA Algorithm with the proposed nonlinear function is briefly introduced in Section 2. We do two Matlab simulations in Section 3 to test and validate the performance and the runtime of the improved Fast-ICA algorithms. The final section is a summary of the content of this paper.

## 2  Methodology

The noise-free model of basic ICA is as follows: $x = As$, where $A$ is a linear mixing matrix. In general, the number of sources is equal to the number of the mixtures. The solution of ICA is to estimate a separation or demixing matrix $B$ as the inverse of $A$, $B = A^{-1}$. Based on the concept of mutual information, we define the differential entropy $H(y)$ of of a random vector $y = R^N$ with the probability density $p(y)$ as follows:

$$H(y) = -\int p(y) \log p(y) dy \tag{1}$$

Negentropy $J(y)$ is defined as follows::

$$J(y) = H(y_G) - H(y) \tag{2}$$

where $y_G$ is a Gaussian random vector of the same covariance matrix as $y$ [7].

To use the definition of ICA given above, an approximation of the negentropy based on the maximum-entropy principle is given by

$$J(y_i) \approx c\{E[G(y_i)] - E[G(v)]\}^2 \tag{3}$$

where $c$ is an irrelevant constant, $E(\cdot)$ is the mean operation, $G(\cdot)$ is practically any non-quadratic contrast function. $v$ is a Gaussian variable with zero mean and unit variance, and the random variable $y_i$ is assumed to be of zero mean and unit variance[9].

The formula (3) gives readily the objective function of the Fast-ICA. To find one independent component (source signal) or projection pursuit direction as $y_i = w^T x$, we maximize the function $J_G$ given by [4]

$$J_G(w) = \{E[G(w^T x)] - E[G(v)]\}^2 \tag{4}$$

where $w$ is a $N$-dimensional (weight) vector constrained so that $E[(w^T x)^2] = 1$. Every vector $w_i (i = 1, 2, ..., N)$ gives one of the rows of the orthogonal matrix $W$ that is derived from ICA[7]. The most frequently used contrast functions $G(\cdot)$ and nonlinear function (nonlinearity) $g(\cdot)$ are summarized as follows [4]:

$$\begin{aligned} G_1(u) &= \log \cosh(u) & g_1(u) &= \tanh(u) \\ G_2(u) &= -e^{-u^2/2} & g_2(u) &= u e^{-u^2/2} \\ G_3(u) &= u^4/4 & g_3(u) &= u^3 \end{aligned} \tag{5}$$

The derivative function $g = G'$ is called the nonlinearity. $g_1(\cdot)$, $g_2(\cdot)$ and $g_3(\cdot)$ are signed as 'tanh','gauss' and 'pow3', respectively.

In practical problems, the computational speed of an algorithm is a factor that limits its applications. Based on this, we devote to finding new nonlinearity. The basic principles of constructing nonlinearities in the Fast-ICA algorithm are as follows: (1) The nonlinearities should be simple and their calculation speed should be fast; (2) The separation performances should be better than that of the traditional nonlinearities.

In this paper, we adopt $g_4(u) = \sin(u)$ as the test nonlinearity because our purpose is to find or design suitable function to replace $g_1, g_2$ and $g_3$ and to accelerate the computation of Fast-ICA and yet achieve the same or better performance. $g_4(\cdot)$ is signed as 'sin'. The improved Fast-ICA Algorithm is listed in Table 1.

Table 1: The Improved FastICA Algorithm

| |
|---|
| **Input:** The number of signal sources $M$. The mixed signals $X$. |
| **Output:** The separation matrix $W$. |
| 1: Zero mean value processing for the mixed signals $X$, and namely $E(X_i) = 0$. |
| 2: Delayed correlation processing, and namely $E(X^T X) = I$. |
| 3: Choose the proposed nonlinear function $g(u) = \sin(u)$. |
| 4: Initialize the number of estimated vectors, Let $p = 1$. |
| 5: Initialize $w_p$, namely $w_p = w_p / \|w_p\|$. |
| 6: Calculate $w_p$ and let $w_p = E\{Xg(w_p^T X)\} - E\{g'(w_p^T X)\}w$. |
| 7: $w_{p+1} = w_{p+1} - \sum_{j=1}^{p} w_{p+1}^T w_j w_j$. |
| 8: If $\|w_{p+1} - w_p\| < \varepsilon$ does not converge, return Step 6. |
| 9: Let $p = p + 1$, if $p < M$, return Step 5. |

## 3 Experiments and Results

In this section, we compare the new nonlinearity with the classical nonlinearities ('tanh', 'gauss' and 'pow3') in the Fast-ICA algorithms by simulation. For each nonlinearity, the experiments are conducted with the same software and hardware environments in order to make the comparison as fair as possible. All the simulations are carried out in MATLAB 2017b, system of the PC is Windows 7(64-bit version), CPU is Intel® Core™ i7-2720QM Processor@2.20GHz, and RAM is 16 GB.

The Fast-ICA algorithm with the four nonlinearities are compared along the two criteria: statistical and computational.

The statistical performance or separation accuracy is measured using a performance index $C$ (correlation coefficient), defined as follows.

$$C(x, y) = \frac{\text{cov}(x, y)}{\sqrt{\text{cov}(x, x)} \sqrt{\text{cov}(y, y)}} \tag{6}$$

$C(x, y) = 0$ means that $x$ and $y$ are uncorrelated, and the signals correlation increases as $C(x, y)$ approaches unity, the signals become fully correlated as $C(x, y)$ becomes unity. The smaller the value $C$ is, the poorer the separation performance of an ICA algorithm.

(2)The computational load is measured using running time needed for convergence[7]. To compare the computational speed, we record the average runtime of algorithm running ten times for different nonlinearities. The mixing square matrix $A$ was randomly generated(such as $A$ = [0.1946 0.8345 0.1477; 0.2252 0.7008 0.2098; 0.0967 0.8110 0.7473]).

In the first experiment, the sources are three music tunes ('guitar.wav', 'piano.wav' and 'trumpet.wav'). The sample rate is 44100. For each source signal, the number of samples N=400000.

In the second experiment, the sources are three commonly used test images('lenna.bmp', 'pepper.bmp' and 'sailboat.bmp'). For each source signal, image size is 512 by 512 and bit depth of image is 24.

The separation results by the improved Fast-ICA algorithm are shown in Fig.1 and Fig.2.

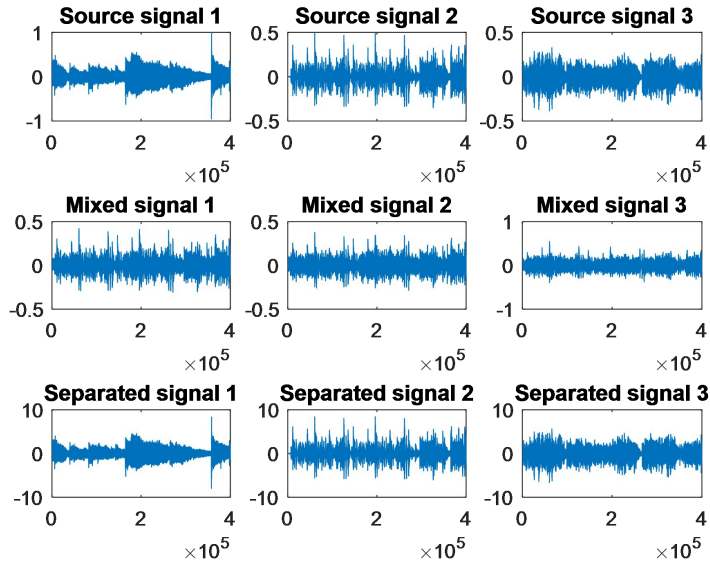

**Fig. 1** *Displaying source signals, mixed signals and separated signals for music - 1, 2 and 3.*

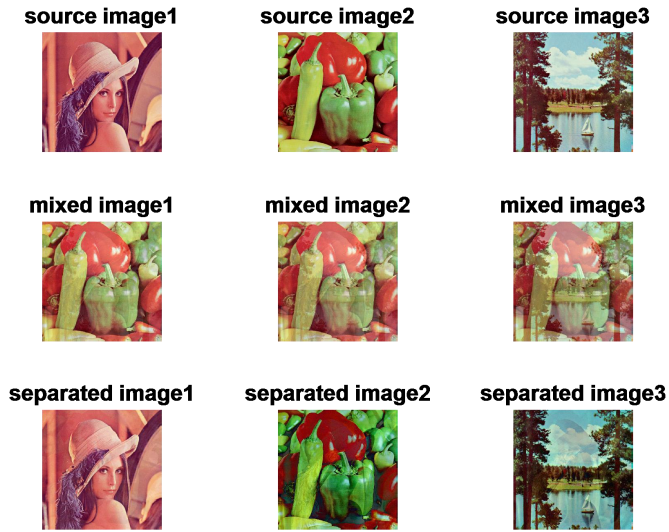

**Fig. 2** *Displaying source images, mixed images and separated images for image - 1, 2 and 3.*

The Fast-ICA algorithms with different nonlinearities are compared with the separation accuracy and the computational speed, the average correlation coefficient (C-ave) and the average runtime(T-ave) after ten executions of the algorithm are listed in Table 2 and Table 3.

**Table 2:** The average correlation coefficient (C-ave) and the average runtime(T-ave) after ten executions of the algorithm based on Experiment 1.

| nonlinearities | C-ave | T-ave(s) |
|---|---|---|
| 'tanh' | 1.0000 | 0.8142 |
| 'gauss' | 1.0000 | 0.8346 |
| 'pow3' | 0.9999 | 1.0158 |
| 'sin' | 1.0000 | 0.5588 |

**Table 3:** The average correlation coefficient (C-ave) and the average runtime(T-ave) after ten executions of the algorithm based on Experiment 2.

| nonlinearities | C-ave | T-ave(s) |
|---|---|---|
| 'tanh' | 0.9621 | 1.6135 |
| 'gauss' | 0.9727 | 1.5860 |
| 'pow3' | 0.9696 | 2.1706 |
| 'sin' | 0.9797 | 1.4326 |

As are shown in Fig. 1 and Fig.2, the sources are well recovered after separation.From the table 2 and 3 , we can see that new nonlinearity 'sin' achieve the same or better separation accuracy and the computational speed is the fastest.

## 4 Conclusions

A novel nonlinearity has been introduced in this paper to be used, there are two advantages of using modified Fast-ICA algorithm. Firstly, the proposed non-linear function 'sin' has the same or better separation accuracy as compared to the commonly used classical functions('tanh', 'gauss' and 'pow3'). Secondly, its computational speed is the fastest. The proposed algorithm has been tested on many different sets of images or audio files. Therefore it is proved that the proposed nonlinearity(Improved Fast-ICA Algorithm) always allows to achieve a better performance of blind source separation technique with a higher execution speed as compared to the commonly used classical functions. The proposed algorithm may be used in other applications of digital signal processing in the future.

References

[1] J. H´erault, C. Jutten, and B. Ans. : 'D´etection de grandeurs primitives dans un message composite par une architecture de calcul neuromim´etique en apprentissage non supervis´e'. Proc. GRETSI, Nice, France, 1985, pp. 1017-1020

[2] Comon P. : 'Independent component analysis，a new concept?', Signal Processing, 1994, 36, (3), pp. 287-314

[3] Hyvärinen A, Oja E.: 'A fast fixed- point algorithm for independent component analysis', Neural Computaion, 1997,9, (7), pp.1483-1492

[4] Hyvärinen A.: 'Fast and robust fixed- point algorithms for independent component analysis', IEEE Transactions on Neural Networks, 1999, 10, (3), pp. 626-634

[5] Dermoune A, Wei T.: 'FastICA algorithm：Five criteria for the optimal choice of the nonlinearity function', IEEE Transactions on Signal Processing, 2013, 61, (8), pp. 2078-2087

[6] Miettinen J, Nordhausen K, Oja H, et al.: 'Deflation-based FastICA with adaptive choices of nonlinearities', IEEE Transactions on Signal Processing, 2014, 62, (21), pp. 5716-5724

[7] He, Xuansen, et al.: 'Large-Scale Super-Gaussian Sources Separation Using Fast-ICA with Rational Nonlinearities', International Journal of Adaptive Control and Signal Processing, 2017, 31, (3), pp. 379-397

[8] Meng, Xin, et al. : 'An Improved FastICA Algorithm Based on Modified-M Estimate Function', Circuits Systems and Signal Processing, 2018, 37, (3), pp. 1134-1144

[9]Hyvarinen, Aapo, et al.: 'Independent Component Analysis', Encyclopedia of Environmetrics, 2001,(1), pp. 151-175